\documentclass{article}
\usepackage{amsmath}
\usepackage{amssymb}
\usepackage{latexsym}
\usepackage{a4wide}
\usepackage[dvips]{graphicx}

\title{Propagating waves pattern in a falling liquid curtain}

\author{N. Le Grand, P. Brunet, L. Lebon and L. Limat}

\date{?? and in revised form ??}
\begin{document}

\maketitle

\rm{$^1$Laboratoire de Physique et M\'ecanique des Milieux H\'et\'erog\`enes, UMR 7636 CNRS, 10, rue Vauquelin 75005 Paris (France)
$^2$ F\'ed\'eration de recherche Mati\`ere et Syst\`emes Complexes, FR CNRS 2438, France}

\begin{abstract}

We have preformed experiments on a liquid curtain falling from a horizontal, wetted, tube and lateraly constrained by two vertical wires. The fluid motion nearly reduces to a free-fall, with a very low detachment velocity below the tube. Thus, the curtain contains a large subsonic area, i.e. a domain where the sinuous waves travel faster than the fluid. The upper boundary not being constrained in the transverse direction, we have observed the appearance of an up to now unreported instability when the flow rate is progressively reduced:  the top of the curtain enters a pendulum-like motion, coupled to a propagative pattern of curtain undulations, structured as a chessboard. Measurements of the phase velocity and frequency of this pattern are reported. Data are in agreement with a simple dimensional argument suggesting that the wave velocity is proportional to the surface tension divided by the mass flux of liquid per unit length. This scaling is also that followed by the fluid velocity at the transonic point, i.e. the point where the fluid velocity equals that of sinuous waves. We finally discuss implications of these results on the global stability of falling curtains.

\end{abstract}

\section{Introduction}

Thin liquid sheet flows are involved in numerous practical applications: curtain coating technics (\cite{Miyamoto97}), atomisation of sheets into droplets (\cite{Lefebvre89}), paper manufacturing (\cite{Alfredsson98}) ... In this context, plane sheets falling under the influence of gravity have been widely studied (\cite{Brown61, Lin81, Lin81(2), Kodak93, deLuca97, Teng97, deLuca99, ecs03}). The first noticeable contribution was due to Brown (\cite{Brown61}), who reported the first experiments and, among other results, showed that the flow nearly reduces to a free fall. He also tried to build a stability criterion by considering the possible evolution of a transient hole through the curtain. Capillary forces tend to increase the hole size, the upper boundary of which is pulled upward, whereas inertia "pushes" it downward.  This led Brown to identify the Weber number $We=\rho h U^2 / 2 \gamma$ built upon surface tension $\gamma$, liquid density $\rho$, local liquid velocity $U$ and local curtain thickness $h$, as a key parameter for the stability. Following his point of view, this number has to be larger than one to guarantee curtain stability. Let us note however that, as the quantity $\Gamma=h U$ (flow-rate per unit length) must be conserved while $U$ increases downward, the Weber number is not uniform, which complicates the possible application of this criterion.

Further studies mainly focussed on surface waves propagation (\cite{Lin81,deLuca97}), by close analogy with previous works from Taylor on axially expanding sheets (\cite{Taylor59}). There are several motivations to this interest: (1) modulations of curtains are  potential sources of imperfection in coating techniques; (2) being sensitive to capillary effects they can be used to measure static or dynamic surface tension (\cite{Lin81}); (3) by analogy with atomisation, decomposition of perturbations upon waves was supposed to give a more natural framework for discussions of curtain stability. One can retain roughly from these studies that two kind of waves can propagate: a symmetrical mode (varicose waves) corresponding to thickness modulations, and an anti-symmetrical mode (sinuous mode) corresponding to modulations of the median transverse position of the curtain (\cite{Taylor59}). These waves have the following velocities (in the limit of an inviscid ambient gas):

\begin{equation}
c_{var} = \sqrt{\frac{\gamma h} {2 \rho}} k
\label{eq:cvar}
\end{equation}

\begin{equation}
c_{sin} = \sqrt{\frac{2 \gamma}{ \rho h}},
\label{eq:csin}
\end{equation}

\noindent respectively for varicose and sinuous waves ($k$ is the wave number). When the fluid velocity $U$ exceeds the largest of these two velocities (in practice the sinuous wave velocity, which often referred to as the most "dangerous" mode), the perturbations are convected downstream and the curtain is supposed to be stable. On the other hand, when $c_{sin} > U$, the waves are able to travel upstream, which suggests that curtain rupture could become possible (provided than an appropriate amplification mechanism can take place). This argument leads exactly to the same result as that proposed earlier by Brown, as far as the curtain stability is concerned, i.e. instability when $We<1$ and stability in the opposite case. This argument has been renewed in terms of absolute and convective instabilities of open flows by several authors (\cite{Lin90,Teng97,deLuca99}), but its relevance is still subject for debates (\cite{Luchini04}).

The practical application of these ideas to curtain stability turns out to be very difficult. Usually, $We$ is not uniform and increases downstream. One has in general a completely stable curtain, or a situation with an upstream unstable domain and a downstream stable region separated by what we will call a "transonic" line (\cite{Brunet04}). The situation is very different from that encountered in atomisation of axially expanding sheets: there is no atomisation front, and the instability becomes a global problem, very sensitive to boundary conditions. This is perhaps why the comparison of theories with experiments (patents) is so deceptive: observance of apparently stable liquid curtains, violating the condition $We>1$ in a large upstream region, are often mentioned (\cite{Kodak93,ecs03}). Also, all the calculations recalled above conclude to the fact that $We$ is the sole relevant number, the viscous effects, for instance, not being clearly taken into account. On the other hand, years of practice in the field of photographic coating (\cite{Miyamoto97}), all conclude to the fact that, for gelatine solutions, there is an optimal viscosity close to 30 times that of water for which the curtain exhibits a maximum of stability. It is to note here that the concept of "stability" is not always very clear, as patents mix the problem of curtain rupture and that of curtain depining from the lateral guides. Nevertheless, it seems that the behaviour of a liquid curtain, including its stability, can hardly be understood from a general point of view. It is necessary to consider each geometry separately with a special care taken with boundary conditions.

\begin{figure}
\begin{center}
(a)\includegraphics[scale=0.6]{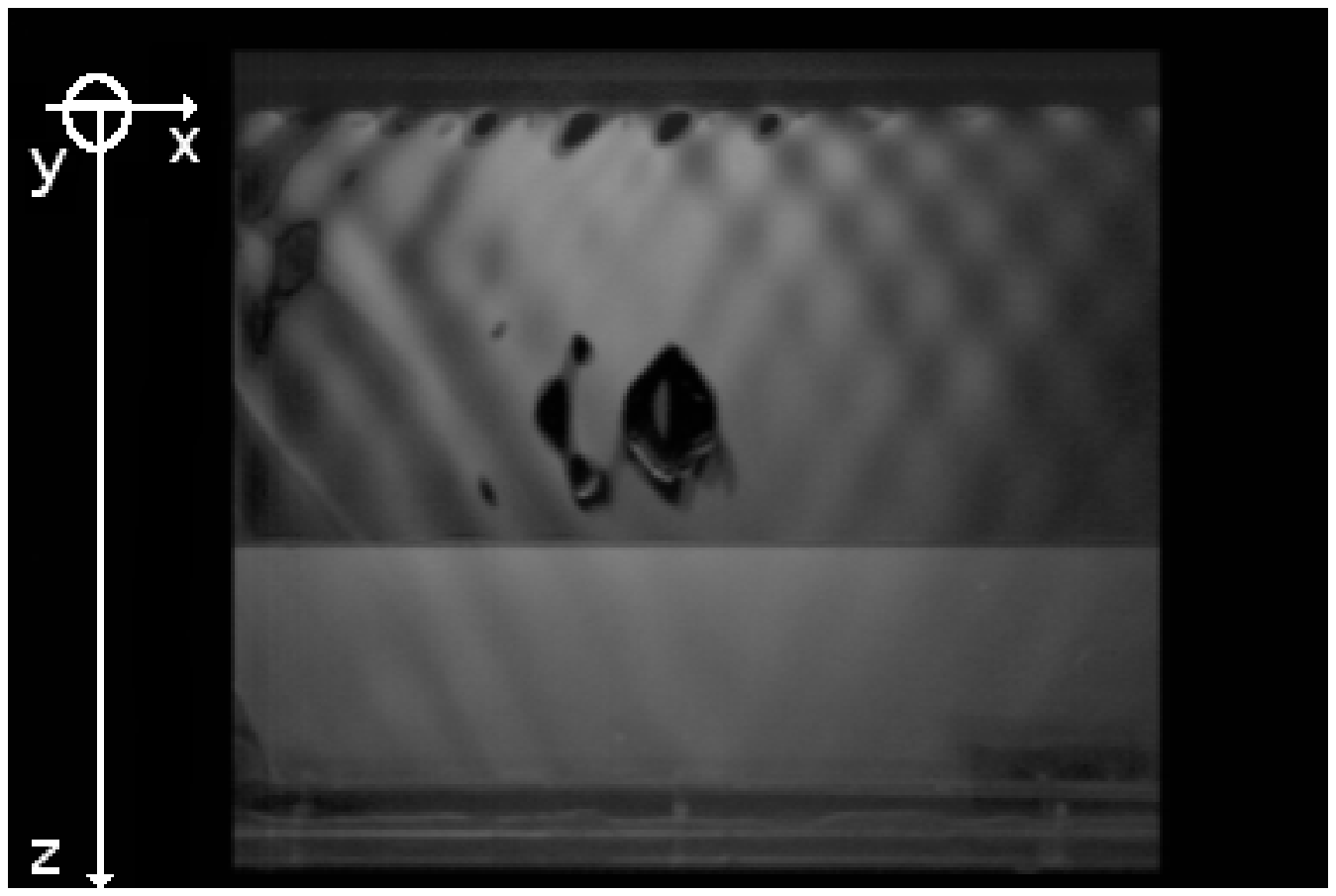}
(b)\includegraphics[scale=0.6]{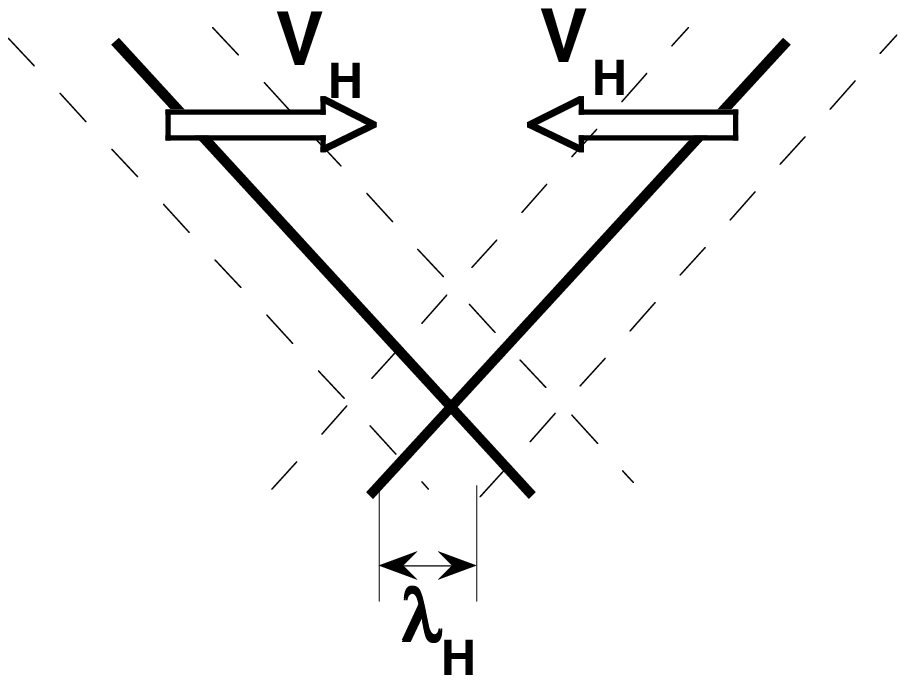}
\caption{(a) Chessboard instability in a liquid curtain. (b) Definition of the horizontal wave velocity $V_H$ and wavelength $\lambda_H$.}
\label{fig:damier}
\end{center}
\end{figure}

In the present paper we investigate experimentally a particular case in which, instead of falling from a slot or from a sharp edge, the liquid is falling from a smoothly curved substrate, in practice a uniformly wetted horizontal tube. This geometry is encountered in several applied coating processes, sometimes combined with a rotation of the tube. We do not solve here the complex related stability problem, but we show that the change of the upper boundary conditions has a dramatic effect on the curtain behaviour: transverse motions of the upper part of the curtain become allowed and, when one reduces the flow rate, a specific instability develops, in which the curtain exhibits a transition towards an oscillatory pattern of waves. This one, reproduced in figure \ref{fig:damier}, is reminiscent of other propagating patterns encountered in free surface instabilities, such as hydrothermal waves or the 1D propagating set of waves formed above a immersed hot wire (\cite{Daviaud97}). We investigate the properties of this up to now unrecognised pattern, in particular its frequency and phase velocity, varying flow-rate, liquid properties and tube radius, and we suggest tentative scaling laws for these quantities. A simple dimensional argument suggests that the phase velocity could be proportional to the critical velocity found at the "transonic" line of the curtain, where the liquid velocity is exactly equal to that of sinuous waves. It seems thus that this instability is linked to intrinsic properties of the curtain and is not only a consequence of unusual boundary conditions. 

Let us mention here that similar observations have been reported very recently in others geometries:  liquid curtain falling across a flat horizontal grid (\cite{Inln04}), and under an overflowing circular dish (\cite{Brunet04}), which shows the generality of this surprising pattern. A brief account of some of our first observations is available in coating congress proceedings (\cite{ecs03}). Though there is nearly no published report on this subject, it turns out that this wavy pattern has also been frequently observed in several industrial coating devices. It constitutes a limit to their efficiency, and also can be used as a precursor signal announcing curtain rupture.

The paper is organized as follows. In section 2, the experimental setup is described. Section 3 presents our observations and data, as well as attempted scaling laws based on dimensional analysis, before the final discussion in section 4. 

\section{Experimental set-up}

\begin{figure}
\begin{center}     
\includegraphics[scale=0.65]{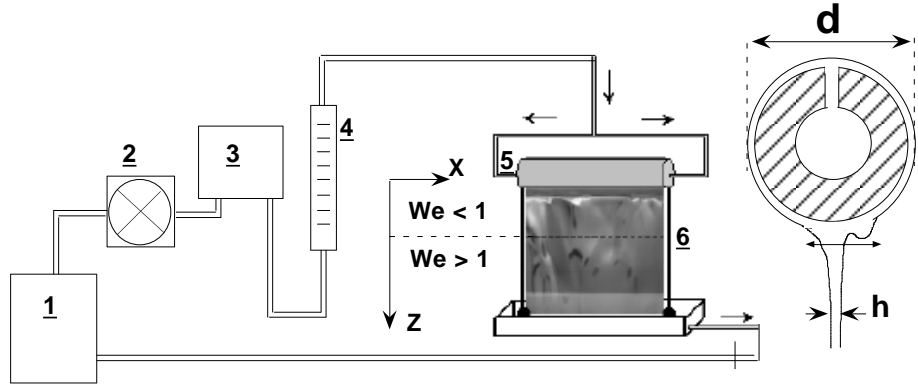}
\caption{Sketch of the experiment. The curtain, guided between two long vertical threads, is falling from a horizontal tube, drilled with a slot turned upward.}
\label{fig:setup}
\end{center}
\end{figure}

The experiment is suggested in figure \ref{fig:setup}. The liquid is pumped from a reservoir (\underline{1}), by means of a  gear-pump \textit{ISMATEC BVP-Z} (\underline{2}) which imposes a constant flow-rate $Q$ measured with a floater flow-meter (\underline{4}). A half-filled chamber (\underline{3}) damps possible residual perturbations. The liquid is then injected at the two ends of a horizontal tube (\underline{5}) (diameter $d$  ranging between 3.4 and 6.8 cm), and flows across a long thin slot (thickness $e$=2 mm) drilled on the upper tube side. If the flow-rate is sufficiently high, a liquid curtain is observed (\underline{6}). Its width of 25.5 cm is kept constant along the vertical direction by two thin nylon threads (diameter 0.01 cm), put under tension by two weights attached at their lower ends. This prevents the curtain shrinkage usually induced  by surface tension. The height of the curtain can be selected at will, ranging between 15 to 25 cm.  All the experiments are performed with silicon oils (Polydimethylsiloxane, PDMS), of different viscosity ranging from 10 to 50 cP.  Their physical properties are given in Table \ref{tab:liq} . The surface tension and density are nearly the same for the three oils.  In the following, the three liquids will be simply referred to as 'V10, 'V30' or 'V50'.  We have checked that the liquid lies at room temperature, which was maintained between 20$^{\circ}$C and 22$^{\circ}$C during the experiments. Special care was taken to protect the system from any sources of perturbations, especially air motions around the experiment. 

\begin{table} 
  \begin{center} 
  \begin{tabular}{lccc} 
      Liquid ref. & Viscosity $\nu$ & Surface tension $\gamma$  & Density $\rho$ \\[3pt] 
       PDMS 47V10 &  10.3 mm$^2$/s  & 20.1 dyn/cm & 0.935 g/cm$^3$ \\ 
       PDMS 47V30 &  32.0 mm$^2$/s & 20.4 dyn/cm & 0.947 g/cm$^3$ \\ 
       PDMS 47V50 &  53.6 mm$^2$/s & 20.7 dyn/cm & 0.957 g/cm$^3$ \\ 
        \label{tab:liq}
        \end{tabular}
  \caption{Physical properties of liquids}
  \end{center} 
\end{table} 

\begin{figure}
\begin{center}     
\includegraphics[scale=0.4]{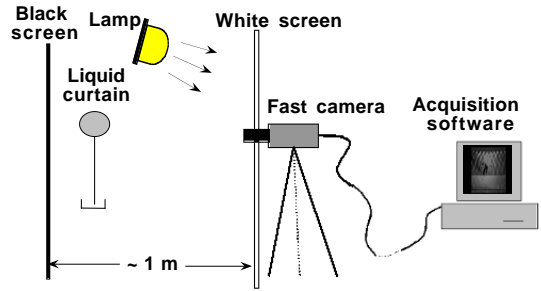}
\caption{Curtain motions visualisation method.}
\label{fig:visurideau}
\end{center}
\end{figure}

Curtain undulations are followed by a high-speed video camera (\textit{FASTCAM} 1024 Motion-Corder). In practice, a frequency of 250 images per second is found to be enough. Because of the required short time acquisition, it is specially important to use powerful, and non-pulsed light sources. The spatial distribution of light had also to be spatially homogeneous. The retained solution, depicted in figure \ref{fig:visurideau}, uses a 300 Watts incandescent lamp which lights a white screen, the latter diffusing a homogeneous light on the curtain. Another screen, black-coloured, is placed behind the curtain, to maximise contrast. A circular hole was made in the white screen, through which the camera lens fits. The reflection of this hole induces a parasite round black shadow on the pictures, of very limited extent, which does not perturb the measurements. The lateral edges are the nylon wires mentioned above, so that the change of boundary conditions as regards usual experiments (\cite{Brown61,Lin81,Kodak93,deLuca99}) concerns the connection of the curtain to the injecting cylinder. The position of the slot has been turned upward in an overflowing configuration (see right insert of figure \ref{fig:setup}). This releases the constraint on the curtain transverse position.

\section{Properties of the chessboard wave pattern}

Experiments are always conducted in the same way: the flow-rate is first increased up to around 5 cm$^2$/s to create the curtain. Its lateral boundaries are put in contact with the vertical nylon wires and then, the flow-rate is progressively decreased. When it reaches a certain threshold, the chessboard wave pattern is observed. A typical example is reproduced on figure  \ref{fig:damier}-a. We observe that  the wave velocity (measured horizontally) is nearly independent on $z$, which leads to the impression of a uniform translation for each set of left and right waves. Figure \ref{fig:sousdamier}-a offers a view from the side and below the injection cylinder, at an orientation of 45$^{\circ}$ with respect to the vertical. This snaking structure is a clear evidence of the sinuous nature of the waves. Under conditions for observance of such pattern, the subsonic area is significantly extended in the curtain, which means that the curtain is highly stretched by gravity. It is worthwhile to give here an insight of the flow in these conditions.

\subsection{Description of the flow}

The flow-rate per unit length $\Gamma$ is simply equal to the total flow-rate $Q$ divided by the curtain width. Typical values for $\Gamma$ are ranging between 0.2 cm$^2$/s and 2 cm$^2$/s. By analogy with the pioneering work of \cite{Brown61}, the velocity should be approximately given by: $U^2 = U_0^2 + 2g \left( z- \alpha (4\nu/\rho)^{2/3} g^{-1/3} \right)$, $\alpha$ being close to unity. In this range of parameters, $U_0$ is smaller than 5 cm/s and this term becomes negligible for $z$ larger than a few millimetres. Also, the offset on $z$ is around 0.3 mm. This was checked by measuring  angles of sinuous wakes below an obstacle (\cite{Lin81,part1}), on the whole curtain. It means that the initial momentum on top of curtain is weak compared to the momentum due to gravity forces. The velocity field takes the simple expression of a free-fall:

\begin{equation}
U^2 = 2gz.
\label{eq:u}
\end{equation}

The origin $z=0$ can be taken at the bottom of the cylinder (figure \ref{fig:damier}-a). The length of the subsonic area is tuned by $\Gamma$, as the position $z^*$ of the vertical coordinate where $We$=1 (which is also the length of the subsonic area), obeys the following condition :  

\begin{equation}
We = \frac{\rho  \Gamma (2 g z^*)^{1/2}}{2 \gamma} = 1,
\label{eq:we}
\end{equation}

which leads to:

\begin{equation}
z^* = \frac{2 \gamma}{g \rho^2 \Gamma^2}
\label{eq:ptrans}
\end{equation}

$We$ increases with $z$, and when $z < z^*$, it is smaller than one. As it modifies the liquid thickness, an increase of $\Gamma$ leads to increase everywhere the local Weber number.

\subsection{Results}

\begin{figure}
\begin{center}
(a)\includegraphics[scale=0.45]{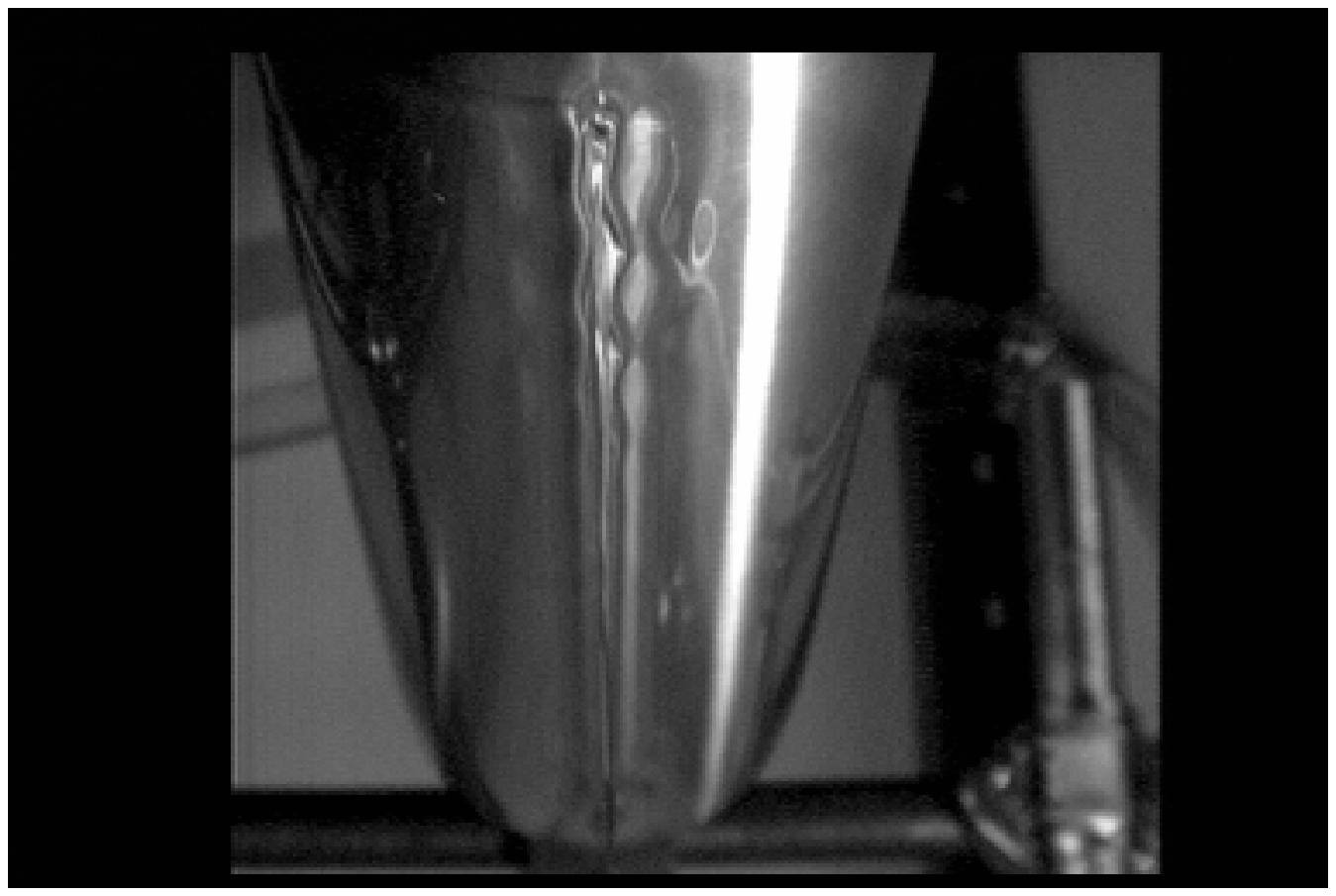}
(b)\includegraphics[scale=0.5]{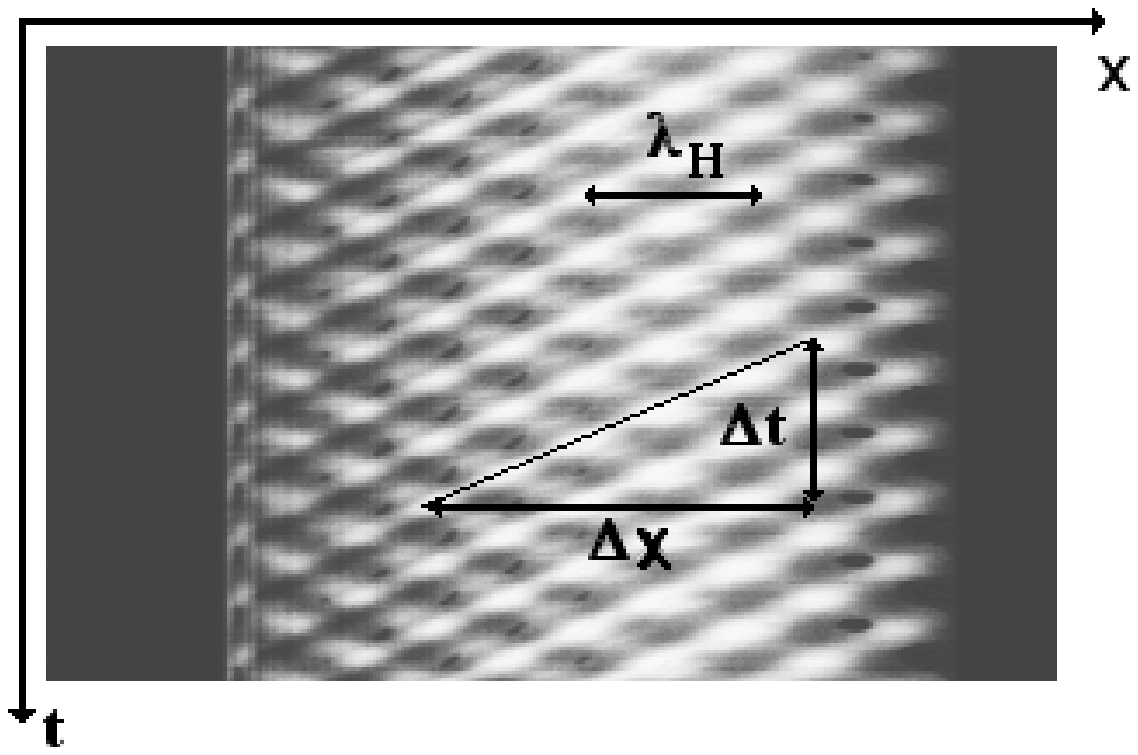}
\caption{(a) View from below of the wave pattern, emphasising its sinuous nature.(b) Spatio-temporal diagram of a chessboard wave pattern.}
\label{fig:sousdamier}
\end{center}
\end{figure}

Measurements of the wave velocity  $V_H = \Delta X / \Delta T$, frequency $f = 1/ \Delta T$, and wavelength $\lambda=\Delta X = V_H / f$ are achieved in the following way: grey levels are extracted along a horizontal line recorded just below the cylinder. By reproducing these grey levels at successive time steps, one creates a spatio-temporal diagram from which $V_H$, $f$ and $\lambda$ can be extracted (figure \ref{fig:sousdamier}-b). Measurements extracted from different lines $z=cte$ on the curtain did not show any variations, but the choice to extract just below the curtain provides the best contrast. 

\begin{figure}
\begin{center}     
(a)\includegraphics[scale=0.49]{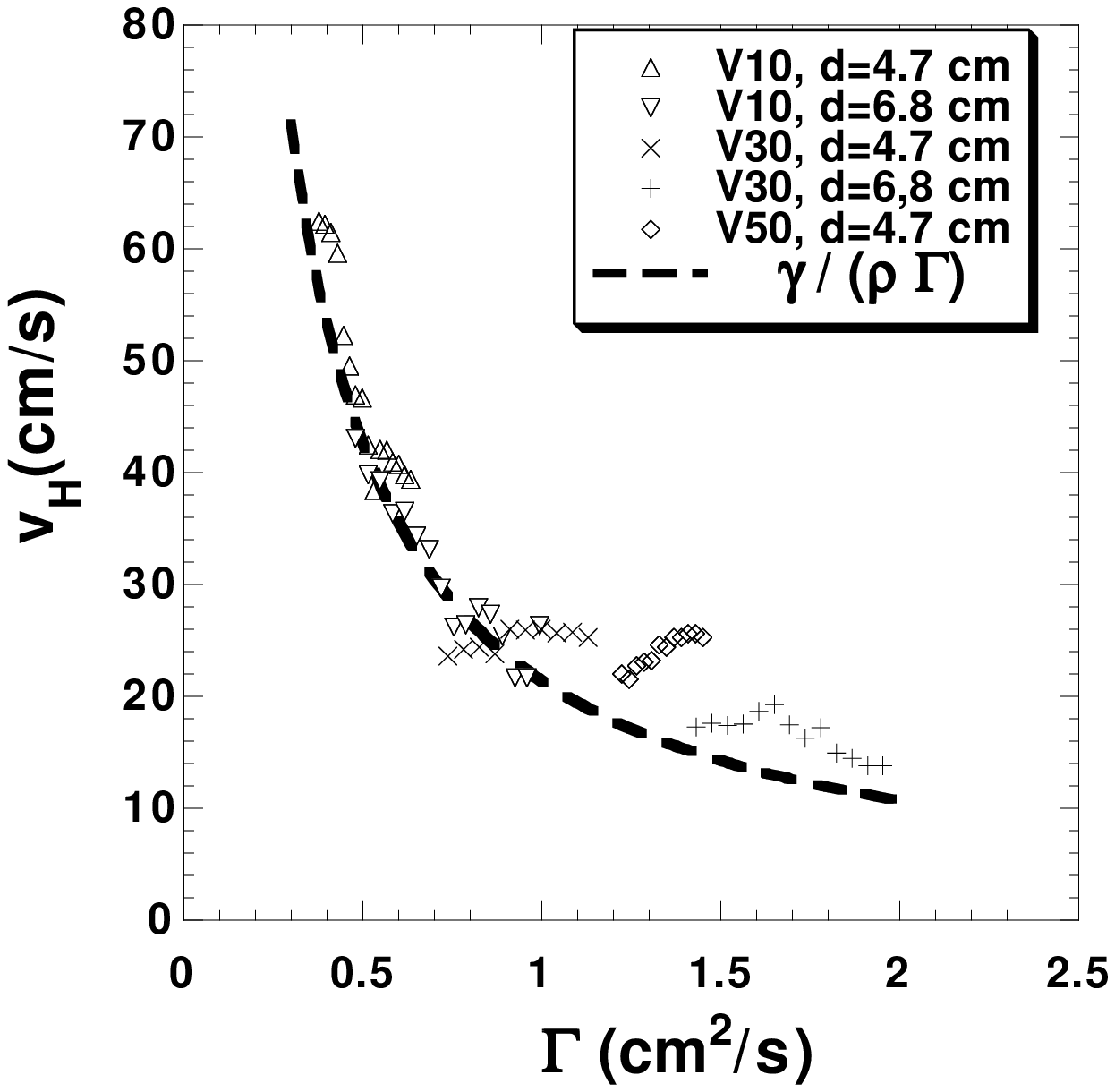}
(b)\includegraphics[scale=0.49]{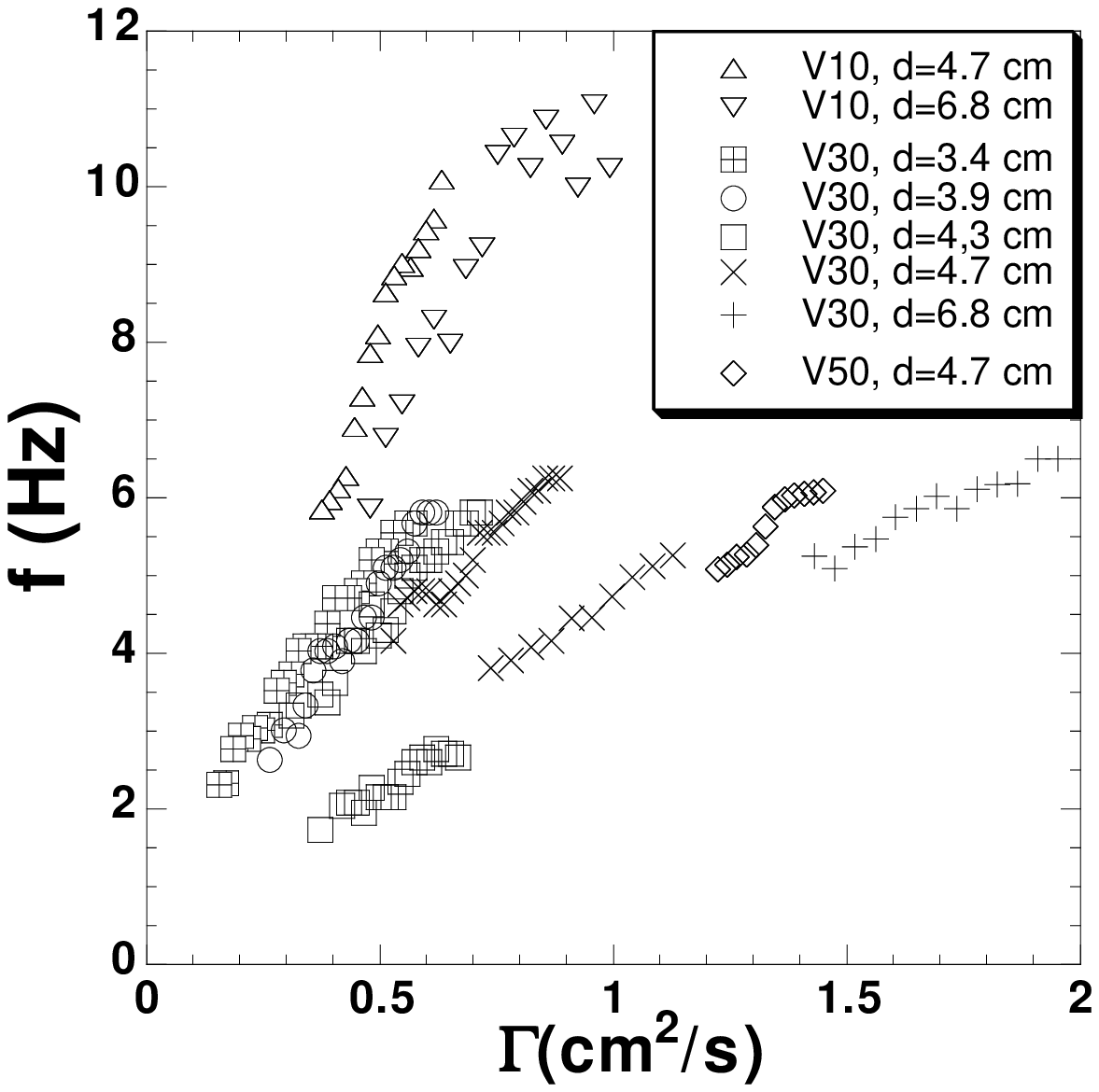}
\caption{(a) Velocity of the travelling waves $V_H$ versus $\Gamma$ for different viscosities and cylinder diameters, superimposed on a characteristic velocity (dashed curve): see text. (b) Frequency versus $\Gamma$.}
\label{fig:damierdiscut}
\end{center}
\end{figure}

Measurements of $V_H$ are plotted on figs. \ref{fig:damierdiscut}-a, versus flow-rate. They concern three viscosities and two cylinder diameters. It turns out that at first order $\nu$ and $d$ do not influence the velocity, whereas they play a role in the range of existence. A strong increase in the speed is noticeable at the smallest flow-rates. The quantity $\gamma / (\rho \Gamma)$ is plotted in dotted line, and it turns out that this quantity fits very well the $V_H$ measurements. Dimensionally, this is the sole speed that can be built with the physical parameters of the system, which does not depend on $z$, nor on $\nu$ and $d$. It is worthwhile to notice that this speed is half the speed of the sinuous waves at the transonic point ($We$=1). Furthermore, taking into account observance of that $V_H$ is nearly independent on $z$, it can be also found dimensionally by combining the two velocities involved in the problem ($c_{sin}$ given by (\ref{eq:csin}) and $U$ given by (\ref{eq:u})). The suitable combination is:

\begin{equation}
\frac{c_{sin}^2}{U} = \frac{2\gamma}{\rho h U}
\label{eq:scalingspeed}
\end{equation}

The frequency is plotted versus $\Gamma$ (figure \ref{fig:damierdiscut}-b). It increases linearly with flow-rate, with a pre-factor depending on viscosity and cylinder diameter. Also, several distinct branches of solutions seem to coexist, which suggests non-trivial non-linear mechanisms. The largest frequencies are obtained for the less viscous liquid (V10), whereas for higher viscosity (V30, V50) data organise in others sets of curves at lower $f$. It is worth noticing that $\lambda_H$ was measured independently from $V_H$ and $f$, and that a relation $\lambda_H \sim \Gamma^{-2}$ was found, which is consistent with $f \sim \Gamma$ and $V_H \sim \Gamma^{-1}$. This scaling is also consistent with (\ref{eq:ptrans}), which suggests again the strong relevance of the transonic line in this problem.

Finally, these plots also contain information on the range of existence of the wave pattern. which depends both on $\nu$ and $d$ in a non-trivial way: a larger $d$ and a larger $\nu$ seems to shift the range to higher flow-rates. For the V50, the range is reduced: this is due to the fact that, when the flow-rate is decreased further, the curtain can separate into two parallel curtains, localised symmetrically to the median plane ($yOz$), or is replaced by a thinner curtain coexisting with an array of columns. For this reason too, it was not possible to observe any wave pattern with the largest diameter $d$=6.8 cm. For high viscosities, the curtain existence range is reduced because of such a separation. This destabilising effect of viscosity may remind the Kapitza instability (\cite{Kapitza}) of a viscous layer of liquid flowing on an inclined plate: destabilising effects due to gravity are more important for thicker layers, which are obtained for higher viscosity (see also \cite{Chang94}).

\begin{figure}
\begin{center}     
\includegraphics[scale=0.55]{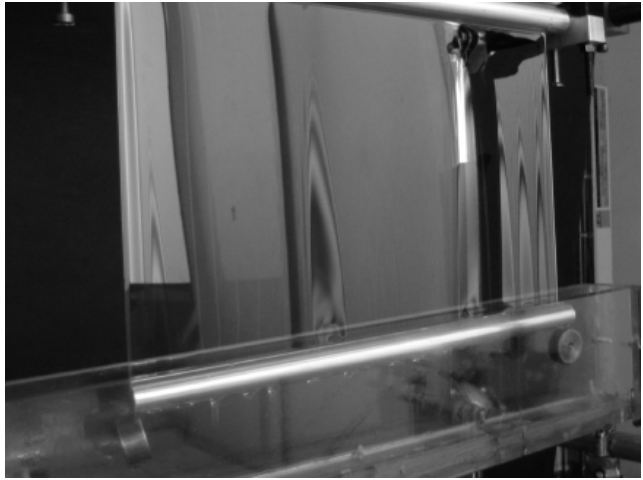}
\caption{An about micron-thick curtain obtained with a bottom boundary condition constrained by a horizontal tube attached to the guides. One can notice several rainbow patterns ($\Gamma$ = 0.015 $cm^2$/s. V30 oil).}
\label{fig:irisations}
\end{center}
\end{figure}

\section{Discussion - Conclusions}

This study reports a new curtain instability leading to a striking pattern of propagative waves. This pattern is presumably due to the pendulum-like oscillations, allowed by the free-constrained boundary conditions on top of the curtain. This may remind some observations on liquid bells formed below an overflowing dish or a porous ring (\cite{Brunet04,Inln04}).

Even if the hydrodynamic mechanisms for velocity and frequency selections still remain unclear, the measurements provide several clues:

- The wave velocity seems to be related to the properties of the transonic line ($z=z^*$): first, its absolute value is half the flow-speed (and so half the sinuous wave-speed) at $z^*$, and is exactly the same on the whole curtain. This is a strong proof for a selection mechanism which involves the transonic point. This quantity also does not depend on $\nu$ nor $d$.

- The frequency follows a linear relationship with flow-rate. However the pre-factor depends on $\nu$, $d$ and presumably on other parameters which influence the complex free-surface shape just below the overhang.

Our observations also question the nature of this pattern in the framework of convective/absolute instabilities in weakly non-parallel flows. Usually (\cite{Huerre93}), a global mode is predicted when the length of the region of absolute instability (here the region $We<$1) is larger than a certain threshold, which can happen when the flow-rate is decreased. The spatial homogeneity of the pattern as well as the velocity selection suggest that it may be identified to such a global mode.

Furthermore, some points of our study could be related to the still disputed problem of curtain break-up. First, it shows that sinuous waves, generally pointed out to cause curtain break-up, can be withstood at relatively high amplitudes without leading to break-up. More generally from our observations, no evidence supports the scenario of break-up from a global mode induced by growth of waves, contrary to recent suggestions (\cite{deLuca99,Lin03}). Also, contrary to the case of a cylindrical jet submitted to the Rayleigh-Plateau instability (where such a break-up scenario is indisputable), surface tension does not promote instability in a liquid sheet, but rather damps it (\cite{Li91}) (even if it increases the velocity of sinuous waves). Then in the presumably unstable part of the curtain where $We<$1(if one trusts into a break-up scenario based on wave amplification), stabilising surface tension effects should dominate inertia and prevent any break-up. This is perhaps related to the discrepancy between linear theories and experimental situations where an entirely subsonic curtain can be observed without break-up (\cite{Kodak93,ecs03}). In every careful experiment that we carried out, break-up was always due to the growth of a hole invading the curtain, and never by amplification of curtain undulations. This is consistent with a recent paper \cite{Luchini04}. In that sense, a globally subsonic liquid curtain can be considered as a 'metastable' object, as it withstands weak perturbations and is broken by stronger ones (\cite{part1}).

This study finally illustrates the influence of the modification of the upper boundary conditions on the dynamics and stability of a falling curtain. Qualitatively, we have also begun to investigate the influence of the bottom boundary condition. A striking result is observed: one can maintain curtains at very low flow-rates by simply adding a cylinder at the bottom of the curtain (\cite{ecs03}). In this situation, hole nucleation at the curtain bottom becomes very difficult, which prevents break-up. Such a situation is shown in figure \ref{fig:irisations}, where rainbow patterns witness the sheet thickness can be locally of the order of a few light wavelengths. Local perturbations on such thin curtains (brought by an obstacle, for example) do not necessary lead to break-up. This surprising effect is presently under study.

\appendix

\end{document}